\documentclass[twoside]{IEEEtran}
\usepackage{graphicx,amsmath,amsbsy,amssymb,citesort,url,subeqn}

\newcommand{\newtext}[1]{#1}

\def\real    { \mathbb{R} }
\def \x {x}
\def \y {y}

\def \j {j} 
\def \J {J} 
\def \n {n} 
\def \N {N} 
\def \numcols {{D'}}
\def \Ebar {E}

\def \Etilde {\overline{E}}
\newcommand \rank[1] {\mathrm{rank}( #1 )}

\newtheorem{THEO}{Theorem}
\newtheorem{LEMM}{Lemma}

\newtheorem{DEFI}{Definition}

\newtheorem{EXAMPLE}{Example}

\newcommand{\qed}{{\unskip\nobreak\hfil\penalty50\hskip2em\vadjust{}
           \nobreak\hfil$\Box$\parfillskip=0pt\finalhyphendemerits=0\par}}

\newcommand{\qq}{\vspace*{0mm}}
\newcommand{\pp}{\vspace*{0mm}}

\begin{document}
\renewcommand{\textfraction}{0}

\title{Measurement Bounds for Sparse Signal Ensembles via Graphical Models}
\IEEEaftertitletext{\vspace{-10mm}\begin{center} {\em This paper is dedicated to the memory of Hyeokho Choi, our colleague, mentor, and friend.}\end{center}\vspace{5mm}}
\author{Marco F.\ Duarte,~\IEEEmembership{Member,~IEEE,} Michael B.\ Wakin,~\IEEEmembership{Member,~IEEE,} 
Dror Baron,~\IEEEmembership{Senior Member,~IEEE,}\\
Shriram Sarvotham,~\IEEEmembership{Member,~IEEE,}  and Richard G.\
Baraniuk,~\IEEEmembership{Fellow,~IEEE,}%
\thanks{This work was supported by the grants NSF CCF-0431150, NSF CCF-0926127, DARPA HR0011-08-1-0078, DARPA/ONR N66001-08-1-2065, ONR N00014-07-1-0936, ONR N00014-08-1-1112, AFOSR FA9550-09-1-0432, AFOSR FA9550-09-1-0465, ARO MURI W911NF-07-1-0185, ARO MURI W911NF-09-1-0383, and the Texas Instruments Leadership University Program. MFD was also supported by NSF Supplemental Funding DMS-0439872 to UCLA-IPAM, P.I.\ R.\ Caflisch. DB was also supported by the Israel Science Foundation while at the Technion -- Israel Institute of Technology. Preliminary versions of this work appeared at the Workshop on Sensor, Signal and Information Processing (SENSIP), 2008~\cite{DCSSensip,SenSIPTR}.}
\thanks{M. F. Duarte is with the Department of Electrical and Computer 
Engineering, University of Massachusetts, Amherst, MA 01003 USA (e-mail: mduarte@ecs.umass.edu)}%
\thanks{M. B. Wakin is with the Department of Electrical Engineering and Computer Science, Colorado School of Mines, Golden, CO 80401 USA (e-mail: mwakin@mines.edu)}%
\thanks{D. Baron is with the Department of Electrical and Computer 
Engineering, North Carolina State University, Raleigh, NC 27695 USA (e-mail: barondror@ncsu.edu)}%
\thanks{S. Sarvotham is with Halliburton Energy Services, Houston, TX 77032 USA (e-mail: shri@rice.edu)}%
\thanks{R. G. Baraniuk is with the Department of Electrical and Computer 
Engineering, Rice University, Houston, TX 77005 USA (e-mail: richb@rice.edu)}
\thanks{Copyright (c) 2013 IEEE. Personal use of this material is permitted. However, permission to use this material for any other purposes must be obtained from the IEEE by sending a request to pubs-permissions@ieee.org.}}
\maketitle \thispagestyle{empty}
\vspace{-0.3in}

\begin{abstract}\noindent
In compressive sensing, a small collection of linear projections of a sparse signal contains enough information to permit signal recovery.
Distributed compressive sensing (DCS) extends this framework by defining ensemble sparsity models, allowing a correlated ensemble of sparse signals to be jointly recovered from a collection of separately acquired compressive measurements.
\newtext{In this paper, we introduce a framework for modeling sparse signal ensembles that quantifies the intra- and inter-signal dependencies within and among the signals}.
This framework is based on a novel bipartite graph representation that links the sparse signal coefficients with the measurements obtained for each signal.
\newtext{Using our framework, we provide fundamental bounds on the number of noiseless measurements that each sensor must collect to ensure that the signals are jointly recoverable.}
\end{abstract}
\normalsize


\section{Introduction}
\label{sec-intro}

A unique framework for signal sensing and compression has recently developed under the rubric of {\em compressive sensing} (CS).
CS builds on the work of Cand\`{e}s, Romberg, and Tao~\cite{CandesRUP} and Donoho~\cite{DonohoCS}, who showed that if a signal $\x \in \real^N$ can be expressed as a sparse superposition of just $K < N$ elements from some dictionary, then it can be recovered from a small number of linear measurements $\y = \Phi \x$, where $\Phi$ is a measurement matrix of size $M \times N$, and $M < N$.
One intriguing aspect of CS is that randomly chosen measurement matrices can be remarkably effective for nonadaptively capturing the information in sparse signals.
In fact, if $x$ is a fixed $K$-sparse signal and just $M = K+1$ random measurements are collected via a matrix $\Phi$ with independent and identically distributed (i.i.d.)\ Gaussian entries, then with probability one $x$ is the unique $K$-sparse solution to $y = \Phi x$~\cite{DCSTR06}.
While there are no tractable recovery algorithms that guarantee recovery when so few measurements are collected, there do exist a variety of practical and provably effective algorithms~\cite{CandesRUP,DonohoCS,TroppOMP,cosamp} that work when $M = O(K \log (N))$.

The current CS theory has been designed mainly to facilitate the sensing and recovery of a single signal $x \in \real^N$.
It is natural to ask whether CS could help alleviate the burdens of acquiring and processing high-dimensional data in applications involving multiple sensors.
Some work to date has answered this question in the affirmative.
For example, if the entries of an unknown vector $x \in \real^N$ are spread among a field of sensors (e.g., if $x$ represents a concatenation of the ambient temperatures recorded by $N$ sensors at a single instant), then certain protocols have been proposed for efficiently computing $y = \Phi x$ through proper coordination of the sensors~\cite{Nowak,Gossiping,Wang,saligrama2}.
Given $y$, standard CS recovery schemes can then be used to recover $x$ using a model for its sparse structure.

It is interesting, however, to consider cases where each sensor observes not a single scalar value but rather a longer vector.
For example, consider an ensemble of signals $x_1, x_2, \dots, x_J \in \real^N$ observed by a collection of $J$ sensors, where each sensor $j \in \{1,2,\dots,J\}$ observes only signal $x_j$ (e.g., $x_j$ might represent a time series of $N$ temperature recordings at sensor position $j$).
In such a scenario, one could employ CS on a sensor-by-sensor basis, recording random measurements $y_j = \Phi_j x_j$ of each signal, and then reconstructing each signal $x_j$ from the measurements $y_j$.
Such an approach would exploit {\em intra}-signal \newtext{dependencies} (manifested in a sparse model for each $x_j$), but would not exploit any {\em inter}-signal \newtext{dependencies} that may exist among the signals $x_j$.

Motivated by this observation, we have proposed a framework known as {\em distributed compressive sensing} (DCS) that allows the exploitation of {\em both} intra- and inter-signal \newtext{dependency} structures.\footnote{\newtext{Our prior work in DCS is contained in a technical report~\cite{DCSTR06} and several conference publications~\cite{DCSAllerton05,DCSAsilomar05,DCSNIPS,DCSIPSN06}.}}
In a typical DCS scenario, each sensor separately collects measurements $y_j = \Phi_j x_j$ as described above, but these measurements are then transmitted to a single collection point (a single ``decoder'') where the ensemble of signals is reconstructed {\em jointly} using a model that characterizes \newtext{dependencies} among the sparse signals.
By exploiting the inter-signal \newtext{dependencies}, DCS allows the overall measurement burden to be shared among the $J$ sensors; in other words, the signal ensemble can be reconstructed jointly from significantly fewer measurements than would be required if each signal were reconstructed individually.
Although we do not go into the details here, one can make interesting connections between DCS and the Slepian-Wolf framework for distributed source coding, in which correlated random sources can each be encoded below their nominal entropy rate if they are decoded jointly~\cite{DCSTR06,SW73,CoverThomas,DCSAllerton05}.

As mentioned above, any DCS decoder must rely on a \newtext{dependency} model that describes the anticipated structure within and among the signals in the ensemble.
There are many conceivable ways in which \newtext{dependencies} can be described among a collection of sparse signals.
We have previously proposed~\cite{DCSTR06,DCSAllerton05,DCSAsilomar05,DCSNIPS,DCSIPSN06} several models for capturing such \newtext{dependencies} and studied each model in isolation, developing a variety of practical reconstruction algorithms and theoretical arguments customized to the nuances of each model.
The goal of this paper is to develop a broader, general purpose framework for quantifying the sparsity of an ensemble of correlated signals.
\newtext{Our analysis framework is rooted in a set of definitions that characterize and quantify the dependencies between signals in the ensemble.}
In Section~\ref{sec:generalframework}, we introduce a factored representation of the signal ensemble that decouples its location information from its value information: a single vector encodes the values of all nonzero signal entries, while a binary matrix maps these values to the appropriate locations in the ensemble.
We term the resulting models {\em ensemble sparsity models} (ESMs).
ESMs are natural and flexible extensions of single-signal sparse models; in fact, the ensemble \newtext{dependency} model proposed in~\cite{dirty} and all of our previously proposed models fit into the ESM framework as special cases.

The bulk of this paper (Section~\ref{sec-theory}, Section~\ref{sec:centralpf}, and several supporting appendices) is dedicated to answering a fundamental question regarding the use of ESMs for DCS: how many measurements must each sensor collect to ensure that a particular signal ensemble is recoverable?
Not surprisingly, this question is much more difficult to answer in the multi-signal case than in the single-signal case.
For this reason, we focus not on tractable recovery algorithms or robustness issues but rather on the foundational limits governing how the measurements can be amortized across the sensors while preserving the information required to uniquely identify the sparse signal ensemble.
To study these issues, we introduce a bipartite graph representation generated from the ESM that reflects, for each measurement, the sparse signal entries on which it depends.
\newtext{This bipartite graph representation allows us to explicitly quantify the degree of dependency within any subset of sparse signals in the ensemble.}
\newtext{We provide new measurement bounds that relate intimately to this quantified dependency.}
While our previous work in DCS has helped inspire a number of algorithms for recovery of real-world signal
ensembles~\cite{jsconst,EldarMMV,HoldEm}, we believe that the results in this paper and the analytical framework that we introduce will help establish a solid foundation for the future development of DCS theory. \newtext{Furthermore, the bounds obtained with our new formulation match those obtained in our prior work, while obviating the need for some of the technical conditions assumed by our prior analysis~\cite{DCSTR06,DCSAllerton05,DCSAsilomar05,DCSNIPS,DCSIPSN06}.}

\qq
\section{Ensemble Sparsity Signal Models} \label{sec-models}
\label{sec:generalframework}
\qq

In this section, we propose a general framework to quantify the sparsity of an ensemble of correlated signals.
Our approach is based on a factored representation of the signal ensemble that decouples its location information from its value information.
Later, in Section~\ref{sec-theory}, we explain how the framework can be used in the joint recovery of sparse signals from compressive measurements, and we describe how such measurements can be allocated among the sensors. 


\subsection{Notation and Definitions}

We use the following notation for signal ensembles.
Let $\Lambda:=\{1,2,\dots,\J\}$ index the $J$ signals in the ensemble.
For a subset $\Gamma \subseteq \Lambda$, we define $\Gamma^C := \Lambda \setminus \Gamma$ as the complement of $\Gamma$.
Denote the signals in the ensemble by $x_\j$, with $\j \in \Lambda$. We assume that each signal $x_\j \in \mathbb{R}^\N$, and we let
$$X = [ x_1^T ~ x_2^T ~ \cdots x_\J^T]^T \in \mathbb{R}^{\J\N}$$
denote the concatenation of the signals.
For a given vector $v$, we use $v(\n)$ to denote the $n^{\mathrm{th}}$ entry of $v$, and we use the $\ell_0$ ``norm'' $\|v\|_0$ to denote the number of nonzero entries in $v$.
Conventionally, $\|v\|_0$ is referred to as the {\em sparsity} of the vector $v$;\footnote{We consider for the sake of illustration---but without loss of generality---signals that are sparse in the canonical basis. All of our analysis can be easily extended to signals that are sparse in any orthonormal basis.} we elaborate on this point below and discuss natural extensions of the concept of sparsity to multi-signal ensembles.

\subsection{Sparse Modeling for a Single Signal}
\label{sec:singlesparse}

To motivate the use of a factored representation for modeling sparsity, we begin by considering the structure of a single sparse signal $x \in \real^N$ that has $K \le N$ nonzero entries.
We note that the degrees of freedom in such a signal are captured in the $K$ locations where the nonzero coefficients occur and in the $K$ nonzero values at these locations.
It is possible to decouple the location information from the value information by writing $x = P \theta$, where $\theta \in \real^K$ contains only the nonzero entries of $x$, and where $P$ is an $N \times K$ identity submatrix\footnote{\newtext{An $N \times K$ {\em identity submatrix}, $K < N$, is a matrix constructed by selecting $K$ columns from the $N \times N$ identity matrix $\mathbf{I}_{N \times N}$. The selected columns need not be adjacent, but their order is preserved from left to right.}} that includes $x$ in its column span.
Any $K$-sparse signal can be written in this manner.

In light of the above, to model the set of all possible sparse signals, define $\mathcal{P}$ to be the set of all identity submatrices of all possible sizes $N \times K'$, with $1 \le K' \le N$.
We refer to $\mathcal{P}$ as a {\em sparsity model}, because the concept of sparsity can in fact be defined within the context of this model.
To be specific, given an arbitrary signal $x \in \real^N$, one can consider all possible factorizations $x = P \theta$ with $P \in \mathcal{P}$.
Among these factorizations, the dimensionality of the unique smallest representation $\theta$ equals the {\em sparsity level} of the signal $x$; in other words, we will have $\mathrm{dim}(\theta) = \|x\|_0$.

\subsection{Sparse Modeling for a Signal Ensemble}

We generalize the formulation of Section~\ref{sec:singlesparse} to the signal ensemble case by considering factorizations of the form
$X = P\Theta$, where $X \in \real^{JN}$ represents the entire signal ensemble as defined above, $P$ is a matrix of size $JN \times Q$ for some integer $Q$, and $\Theta \in \real^Q$.
In any such factorization, we refer to $P$ and $\Theta$ as the {\em location matrix} and {\em value vector}, respectively.
\begin{DEFI}
An {\em ensemble sparsity model} (ESM) is a set $\mathcal{P}$ of admissible location matrices $P$. The number of columns among the $P \in \mathcal{P}$ may vary, but each has $JN$ rows.
\end{DEFI}
As we discuss further below, there are a number of natural choices for what should constitute a valid location matrix $P$, and consequently, there are a number of possible ESMs that could be used to describe the \newtext{dependencies} among sparse signals in an ensemble.

For a fixed ESM, not every matrix $P \in \mathcal{P}$ can be used to generate a given signal ensemble $X$.
\begin{DEFI}
\label{def:joint_spars} 
\newtext{For a given ensemble $X$ and ESM $\mathcal{P}$, the set of {\em feasible location matrices} is}
\begin{equation*}
\newtext{\mathcal{P}_F(X) := \{P \in \mathcal{P}~\textrm{s.t.}~X \in \mathrm{colspan}(P)\},}
\end{equation*}
\newtext{where $\mathrm{colspan}(P)$ denotes the column span of $P$. In the context of an ESM $\mathcal{P}$, the {\em ensemble sparsity level} of a signal ensemble $X$ is}
\begin{equation*}
\newtext{D = D(X,\mathcal{P}) := \min_{P \in \mathcal{P}_F(X)} \mathrm{dim}(\mathrm{colspan}(P)).}
\end{equation*}
\end{DEFI}
Note that $\mathcal{P}_F(X) \subseteq \mathcal{P}$.
When $P$ is full rank, the dimension of its column span is equal to its number of columns; we will expand on this property in Section~\ref{sec:ciesm}.
For many ESMs, we may expect to have $D < \sum_{j \in \Lambda} \| x_j \|_0$.

\subsection{Common/Innovation Location Matrices}
\label{sec:cilm}

There are a number of natural choices for the location matrices $P$ that could be considered for sparse modeling of a signal ensemble.
In this paper (as we studied earlier in~\cite{DCSTR06}), we are interested in the types of multi-signal \newtext{dependencies} that arise when a number of sensors observe a common phenomenon (which may have a sparse description) and each of those same sensors observes a local innovation (each of which may also have a sparse description).
To support the analysis of such scenarios, we restrict our attention in this paper to location matrices $P$ of the form
\begin{equation}
P = \left[
\begin{array}{ccccc}
P_C & P_1 & {\bf 0} & \hdots & {\bf 0}\\
P_C & {\bf 0} & P_2 & \hdots & {\bf 0}\\
\vdots & \vdots &\vdots&\ddots&\vdots\\
P_C & {\bf 0}& {\bf 0} &\hdots& P_J \end{array} \right], \label{eq:pmatrix}
\end{equation}
where $P_C$ and each $P_j$, $j \in \Lambda$, are identity submatrices with $N$ rows, and where each {\bf 0} denotes a matrix of appropriate size with all entries
equal to 0.
For a given matrix $P$ of this form, let $K_C(P)$ denote the number of columns of the element $P_C$ contained in $P$, and for each $j \in \Lambda$, let $K_j(P)$ denote the number of columns of $P_j$.

Let us explain why such location matrices are conducive to the analysis of signals sharing the common/innovation structure mentioned above.
When a signal ensemble $X \in \real^{JN}$ is expressed as $X = P \Theta$ for some $P$ of the form (\ref{eq:pmatrix}), we may partition $\Theta$ into the corresponding components
\begin{equation*}
\Theta = [\theta_C^T~\theta_1^T~\theta_2^T~\ldots~\theta_J^T]^T,
\end{equation*}
where $\theta_C \in \real^{K_C(P)}$ and each $\theta_j \in \real^{K_j(P)}$. Then, letting
\begin{equation}
z_C := P_C \theta_C ~~~\mathrm{and}~~~ z_j := P_j \theta_j ~\mathrm{for~each}~ j \in \Lambda,
\label{eq:zczj}
\end{equation}
we can write each signal in the ensemble as 
$$x_j = z_C + z_j,$$ 
where the {\em common component} $z_C$ has sparsity $K_C(P)$ and is present in each signal and the {\em innovation components} $z_1, z_2, \dots z_J$ have sparsities $K_1(P), K_2(P), \dots, K_J(P)$, respectively, and are unique to the individual signals.
%

\begin{EXAMPLE}
Consider $J=2$ signals of dimension $N=4$ each, specifically $x_1 = [3~1~0~0]^T$ and $x_2 = [1~1~0~0]^T$.
Different choices of $P$  can account for the common structure in $x_1$ and $x_2$ in different ways.
For example, we could take
%
\begin{equation}
P_C = \left[
\begin{array}{ccc}
1 & 0 \\
0 & 1 \\
0 & 0 \\
0 & 0 \\
\end{array} \right],~~P_1 = \left[
\begin{array}{ccc}
1 \\
0 \\
0 \\
0 \\
\end{array} \right],~\textrm{and}~~P_2 = [\;\;],
\label{eq:p1}
\end{equation}
in which case we can write $X = P\Theta$ by taking $\Theta = [1~1~2]^T$. Under this choice of $P$, we have $z_C = [1~1~0~0]^T$, $z_1 = [2~0~0~0]^T$ and $z_2 = [0~0~0~0]^T$, and the sparsity levels for the respective components are $K_C(P) = 2$, $K_1(P) = 1$, and $K_2(P) = 0$.
Alternatively, we could take
\begin{equation}
\widetilde{P}_C = P_C, ~\widetilde{P}_1 = P_2,~\textrm{and}~\widetilde{P}_2 = P_1,
\label{eq:p2}
\end{equation}
in which case we can write $X = \widetilde{P}\widetilde{\Theta}$ by taking $\widetilde{\Theta} = [3~1~-2]^T$. Under this choice of $\widetilde{P}$, we have $\widetilde{z}_C = [3~1~0~0]^T$, $\widetilde{z}_1 = [0~0~0~0]^T$ and $\widetilde{z}_2 = [-2~0~0~0]^T$, and the sparsity levels for the respective components are ${K}_C(\widetilde{P}) = 2$, ${K}_1(\widetilde{P}) = 0$, and ${K}_2(\widetilde{P}) = 1$.
\label{ex:ex1}
\end{EXAMPLE}

\subsection{Common/Innovation ESMs}
\label{sec:ciesm}

In this paper, we restrict our attention to ESMs that are populated only with a selection of the common/innovation location matrices described in Section~\ref{sec:cilm}.
\begin{DEFI}
An ESM $\mathcal{P}$ is called a {\em common/innovation ESM} if every $P \in \mathcal{P}$ has the form (\ref{eq:pmatrix}) and is full-rank.
\end{DEFI}
The requirement that each $P \in \mathcal{P}$ have full rank forbids any $P$ for which $P_C$ and all $\{P_j\}_{j \in \Lambda}$ have one or more columns in common;
%
%
it is natural to omit such matrices, since a full-rank matrix of the form (\ref{eq:pmatrix}) could always be constructed with equivalent column span by removing each shared column from $P_C$ or any one of the $P_j$.

Depending on the type of structure one wishes to characterize within an ensemble, a common/innovation ESM $\mathcal{P}$ could be populated in various ways. For example:
\begin{itemize}
\item One could allow $\mathcal{P}$ to contain {\em all} full-rank matrices $P$ of the form (\ref{eq:pmatrix}). This invokes a sparse model for both the common and innovation components.
\item Or, one could consider only full-rank matrices $P$ of the form (\ref{eq:pmatrix}) where $P_C = \mathbf{I}_{N \times N}$. This removes the assumption that the common component is sparse.
\item Alternatively, one could consider only full-rank matrices $P$ of the form (\ref{eq:pmatrix}) where $P_C = [\;\;]$ and where $P_1 = P_2 = \cdots = P_J$. This model assumes that no common component is present, but that the innovation components all share the same sparse support.
\item Finally, one could consider only full-rank matrices $P$ of the form (\ref{eq:pmatrix}) where $P_C = [\;\;]$ and where all of the matrices $P_j$ share some minimum number of columns in common. This model assumes that all innovations components share some support indices in common.
\end{itemize}
We have previously studied each of the first three cases above~\cite{DCSTR06,DCSAllerton05,DCSAsilomar05,DCSNIPS,DCSIPSN06}, proposing a variety of practical reconstruction algorithms and theoretical arguments customized to the nuances of each model.
Later, the fourth case above was proposed and studied in~\cite{dirty}.
In this paper, however, we present a unified formulation, treating each model as a special case of the more general common/innovation ESM framework.
Consequently, the theoretical foundation that we develop starting in Section~\ref{sec-theory} is agnostic to the choice of {\em which} matrices $P$ of the form (\ref{eq:pmatrix}) are chosen to populate a given ESM $\mathcal{P}$ under consideration, and therefore our results apply to all of the cases in~\cite{DCSTR06,DCSAllerton05,DCSAsilomar05,DCSNIPS,DCSIPSN06,dirty}.

\qq
\section{Distributed Measurement Bounds} \qq
\label{sec-theory}

In this section, we present our main results concerning the measurement and reconstruction of signal ensembles in the context of common/innovation ESMs.

\subsection{Distributed Measurements}

We focus on the situation where distributed measurements of the signals in an ensemble $X \in \mathbb{R}^{\J\N}$ are collected.
More precisely, for each $j \in \Lambda$, let $\Phi_j$ denote a measurement matrix of size $M_j \times N$, and let $y_j = \Phi_j x_j$ represent the measurements collected of component signal $x_j$.
When appropriate below, we make explicit an assumption that the matrices $\Phi_j$ are drawn randomly with i.i.d.\ Gaussian entries, though other random distributions could also be considered.

Suppose that the collection of measurements
$$
Y = [ y_1^T ~ y_2^T ~ \cdots y_\J^T]^T \in \mathbb{R}^{\sum_{j=1}^J M_j}
$$
is transmitted to some central node for reconstruction.
Defining
\begin{equation*}
\Phi = \left[
\begin{array}{cccc}
\Phi_1 & {\bf 0} & \hdots & {\bf 0}\\
 {\bf 0} & \Phi_2 & \hdots & {\bf 0}\\
\vdots &\vdots&\ddots&\vdots\\
 {\bf 0}&{\bf 0}&\hdots& \Phi_J \end{array} \right] \in \mathbb{R}^{(\sum_{j=1}^J M_j) \times JN},
\end{equation*}
we may write $Y = \Phi X$.
In the context of a common/innovation ESM $\mathcal{P}$, we are interested in characterizing the requisite numbers of measurements $M_1, M_2, \dots, M_J$ that will permit the decoder to perfectly reconstruct the ensemble $X$ from $Y$.

\subsection{Reconstruction of a Value Vector}

Let us begin by considering the case where the decoder has knowledge of some full-rank location matrix $P \in \mathcal{P}_F(X)$.
In this case, perfect reconstruction of the ensemble $X$ is possible if the decoder can identify the unique value vector $\Theta$ such that $X = P \Theta$.
%
%

To understand when perfect reconstruction may be possible, note that for any $\Theta$ such that $X = P \Theta$, we can write
\begin{align}
Y &= \Phi X = \Phi P \Theta \label{eq:pphi}\\
&=
\underbrace{
\left[
\begin{array}{ccccc}
\Phi_1 P_C & \Phi_1 P_1 & {\bf 0} & \hdots & {\bf 0}\\
\Phi_2 P_C & {\bf 0} & \Phi_2 P_2 & \hdots & {\bf 0}\\
\vdots & \vdots &\vdots&\ddots&\vdots\\
\Phi_J P_C & {\bf 0}& {\bf 0} &\hdots& \Phi_J P_J \end{array} \right]}_{\Upsilon} \underbrace{\left[ \begin{array}{c} \theta_C \\ \theta_1 \\ \theta_2 \\ \vdots \\ \theta_J \end{array} \right]}_{\Theta}. \nonumber
\end{align}
To ensure that $\Theta$ can be uniquely recovered from $Y$, certain conditions must be met.
For example, it is clear that the total number of measurements cannot be smaller than the total number of unknowns, i.e., that we must have
\begin{equation}
\sum_{j=1}^J M_j \; \geq \; \mathrm{dim}(\Theta) \; = \; K_C(P) + \sum_{j=1}^J K_j(P).
\label{eq:ncond1}
\end{equation}
However, only certain distributions of these measurements among the sensors will actually permit recovery.
For example, the component  $\theta_j \in \mathbb{R}^{K_j(P)}$ is measured only by sensor $j$, and so we require that
\begin{equation}
M_j \geq K_j(P)
\label{eq:ncond2}
\end{equation}
for each $j \in \Lambda$.
Taken together, conditions (\ref{eq:ncond1}) and (\ref{eq:ncond2}) state that each sensor must collect enough measurements to allow for recovery of the local innovation component, while the sensors collectively must acquire at least $K_C(P)$ extra measurements to permit recovery of the common component.
While these conditions are indeed necessary for permitting recovery of $\Theta$ from $Y$ (see Theorem~\ref{theo:cnv}), they are not sufficient---there are additional restrictions governing how these extra measurements may be allocated to permit recovery of the common component.

To appreciate the reason for these additional restrictions, consider the case where for some indices $n \in \{1,2,\dots,N\}$ and $j \in \Lambda$, row $n$ of $P_C$ contains a $1$ and row $n$ of $P_j$ contains a $1$.
Recalling the definitions of $z_C$ and $z_j$ from (\ref{eq:zczj}), this implies that both $z_C(n)$ and $z_j(n)$ have a corresponding entry in the unknown value vector $\Theta$.
In such an event, however, it is impossible to recover the values of both $z_C(n)$ and $z_j(n)$ from measurements of $x_j$ alone because these pieces of information are added into the single element $x_j(n) = z_C(n) + z_j(n)$.
Intuitively, since the correct value for $z_j(n)$ can only be inferred from $y_j$, it seems that the value $z_C(n)$ can only be inferred using measurements of other signals that do not feature the same overlap, i.e., from those $y_{j'}$ such that row $n$ of $P_{j'}$ contains all zeros.

Based on the considerations above, we propose the following definition.
%
\begin{DEFI}
For a given location matrix $P$ belonging to a common/innovation ESM $\mathcal{P}$ and a given set of signals $\Gamma \subseteq \Lambda$, the {\em overlap size} $K_C(\Gamma,P)$ is the number of indices in which there is overlap between the common and innovation component supports at all signals
$j \in \Gamma^C$:
\begin{align}
\small
K_C(\Gamma,P) :=& \left|\{n \in \{1,\ldots,N\} : \mathrm{~row~} n \mathrm{~of~} P_C\mathrm{~is~nonzero} \right. \nonumber \\
& \left.\mathrm{and~} \forall j \notin \Gamma, \mathrm{~row~} n \mathrm{~of~} P_j
\mathrm{~is~nonzero} \} \right|.
\label{eq:kintersect}
\end{align}
We note that $K_C(\Lambda,P) = K_C(P)$ and $K_C(\emptyset,P) = 0$.
\end{DEFI}


Relating to our discussion above, for each entry $n \in \{1,\ldots,N\}$ counted in $K_C(\Gamma,P)$, we expect that some sensor in $\Gamma$ must take one extra measurement to account for that entry of the common component---it is impossible to recover such entries from measurements made only by sensors outside $\Gamma$.
Our first main result confirms that ensuring the sensors in every $\Gamma \subseteq \Lambda$ collectively acquire at least $K_C(\Gamma,P)$ extra measurements is indeed sufficient to permit recovery of $\Theta$ from $Y$.

\begin{THEO}
{\em (Achievable, known $P$)} Let $X$ denote a signal ensemble, and let $P \in \mathcal{P}_F(X)$ be a full-rank location matrix in a common/innovation ESM $\mathcal{P}$.
For each $j \in \Lambda$, let $\Phi_j$ be a random $M_j \times N$ matrix populated with i.i.d.\ Gaussian entries.
If
\begin{equation}
\sum_{j \in \Gamma} M_j \; \ge \; \left( \sum_{j \in \Gamma} K_j(P) \right) + K_C(\Gamma,P)
\label{eq:achCondition1}
\end{equation}
for all subsets $\Gamma \subseteq \Lambda$, then with probability one over $\{\Phi_j\}_{j \in \Lambda}$,
there exists a unique
solution $\widehat{\Theta}$ to the system of equations $Y = \Phi P
\widehat{\Theta}$, and hence, letting $\widehat{X} := P \widehat{\Theta}$ we have $\widehat{X} = X$.
\label{theo:achstep1}
\end{THEO}

Our proof of Theorem~\ref{theo:achstep1} is presented in Section~\ref{sec:centralpf}.
The proof is based on a bipartite graph formulation that represents the dependencies between the obtained measurements $Y$ and the coefficients in the value vector $\Theta$.
Intuitively, the bipartite graph arises from an interpretation of the matrix $\Upsilon = \Phi P$ as a biadjacency matrix~\cite{berge}.
The graph is fundamental both in the derivation of the number of measurements needed for each sensor and in the formulation of a combinatorial recovery procedure for the case where $P$ is unknown; we revisit that problem in Section~\ref{sec:unk} below.

Although Theorem~\ref{theo:achstep1} can be invoked with any feasible location matrix,  \newtext{the right hand side of (\ref{eq:achCondition1}) takes its lowest values when Theorem~\ref{theo:achstep1} is} invoked using a location matrix that contains just $D$ columns.
One implication of this theorem is that, when a location matrix $P \in \mathcal{P}_F(X)$ is known, reconstruction of a signal ensemble $X$ can be achieved using fewer than $\|x_j\|_0$ measurements at some or all of the sensors $j$.
This highlights the benefit of joint reconstruction in DCS.

Our second main result establishes that the the measurement bound presented in Theorem~\ref{theo:achstep1} cannot be improved.
We defer the proof of the following theorem to Appendix~\ref{app:cnv}.

\begin{THEO}
{\em (Converse)} Let $X$ denote a signal ensemble, and let $P \in \mathcal{P}_F(X)$ be a full-rank location matrix in a common/innovation ESM $\mathcal{P}$.
%
%
For each $j \in \Lambda$, let $\Phi_j$ be an $M_j \times N$ matrix (not necessarily random).
If
\begin{equation}
\sum_{j \in \Gamma} M_j \; < \;  \left( \sum_{j \in \Gamma} K_j(P) \right) + K_C(\Gamma,P)
\label{eq:cnvCondition1}
\end{equation}
for some nonempty subset $\Gamma \subseteq \Lambda$, then there exists a value vector $\widehat{\Theta}$ such that $Y = \Phi P \widehat{\Theta}$ but $\widehat{X} := P \widehat{\Theta} \neq X$.
\label{theo:cnv}
\end{THEO}

\begin{EXAMPLE}
Consider again the signal ensemble presented in Example~\ref{ex:ex1}.
For the matrix $P$ specified in (\ref{eq:p1}), the overlap sizes are $K_C(\{1\},P) = 0$ (since there is no overlap between common and innovation components in sensor 2), $K_C(\{2\},P) = 1$ (since there is overlap in the common and innovation components at sensor 1 for index 1), and $K_C(\{1,2\},P) = K_C(P) = 2$.
Alternatively, for the matrix $\widetilde{P}$ specified in (\ref{eq:p2}), the overlap sizes are ${K}_C(\{1\},\widetilde{P}) = 1$, ${K}_C(\{2\},\widetilde{P}) = 0$, and ${K}_C(\{1,2\},\widetilde{P}) = {K}_C(\widetilde{P}) = 2$.
Thus, for a decoder with knowledge of either one of these location matrices, Theorem~\ref{theo:achstep1} tells us that $X$ can be uniquely recovered if $M_1 \ge 1$, $M_2 \ge 1$, and $M_1 + M_2 \ge 3$.
Conversely, Theorem~\ref{theo:cnv} tells us that $X$ cannot be uniquely recovered using either of these location matrices if $M_1 = 0$, if $M_2 = 0$, or if $M_1 = M_2 = 1$.
\label{ex:ex3}
\end{EXAMPLE}

\qq
\subsection{Identification of a Feasible Location Matrix}
\label{sec:unk}
\qq

In general, when presented with only the measurements $Y$, it may be necessary for a decoder to find {\em both} a feasible location matrix $P \in \mathcal{P}_F(X)$ and a value vector $\Theta$ such that $X = P \Theta$.
Just as identifying the sparse coefficient locations in single-signal CS can require more measurements than solving for the values if the locations are known~\cite{DCSTR06}, the multi-signal problem of jointly recovering $P$ and $\Theta$ could require more measurements than specified in Theorem~\ref{theo:achstep1} for the case where $P$ is known.
Our final main result, however, guarantees that a moderate increase in the number of measurements beyond the bound specified in~(\ref{eq:achCondition1}) is sufficient.
\begin{THEO}
{\em (Achievable, unknown $P$)} Let $X$ denote a signal ensemble, and let $\mathcal{P}$ denote a common/innovation ESM.
For each $j \in \Lambda$, let $\Phi_j$ be a random $M_j \times N$ matrix populated with i.i.d.\ Gaussian entries.
If there exists a full-rank location matrix $P^* \in \mathcal{P}_F(X)$ such that
\pp
\begin{equation}
\sum_{j \in \Gamma} M_j \; \ge \; \left(  \sum_{j \in \Gamma} K_j(P^*) \right) + K_C(\Gamma,P^*) + |\Gamma|
\label{eq:dwtb_loose_achievable_condition_13}
\end{equation}
for all subsets $\Gamma \subseteq \Lambda$, then $X$ can be
recovered from $Y$.
\label{theo:achstep2}
\end{THEO}

\newtext{The achievable measurement bound in Theorem~\ref{theo:achstep2} can be met by taking just one additional measurement per sensor above the rate specified in Theorem~\ref{theo:achstep1}. This additional measurement per sensor is used to cross-validate each possible location matrix $P \in \mathcal{P}$; see the proof in Appendix~\ref{app:achstep2} for details. Note that Theorem~\ref{theo:cnv} also serves as a converse for Theorem~\ref{theo:achstep2}.}
Like Theorem~\ref{theo:achstep1}, Theorem~\ref{theo:achstep2} yields the \newtext{lowest} bounds when invoked using a location matrix that contains just $D$ columns.


The proof of Theorem~\ref{theo:achstep2} involves an algorithm based on an enumerative search over all $P \in \mathcal{P}$; this is akin to the $\ell_0$ minimization problem in single-signal CS.
Indeed, removing the common component and taking $J=1$, our bound reduces to the classical single-signal CS result that $K+1$ Gaussian random measurements suffice with probability one to enable recovery of a fixed $K$-sparse signal via $\ell_0$ minimization~\cite{VenBres98,DCSTR06}.
Although such an algorithm may not be practically implementable or robust to measurement noise, we believe that our Theorem~\ref{theo:achstep2} (taken together with Theorems~\ref{theo:achstep1} and~\ref{theo:cnv}) provides a theoretical foundation for understanding the core issues surrounding the measurement and reconstruction of signal ensembles in the context of ESMs.

\newtext{It is worth noting that the bounds obtained in Theorems~\ref{theo:achstep1}, \ref{theo:cnv}, and \ref{theo:achstep2} match those we previously obtained for specific signal ensemble dependencies~\cite{DCSTR06,DCSAllerton05,DCSAsilomar05,DCSNIPS}. However, in contrast to our prior results, Theorems~\ref{theo:achstep1}--\ref{theo:achstep2} do not require a statistical model for the signal ensemble~\cite{DCSAllerton05,DCSAsilomar05,DCSNIPS} and are not asymptotic in the number of signals in the ensemble~\cite{DCSAsilomar05,DCSNIPS}.}

\begin{EXAMPLE}
We once again revisit the signal ensemble presented in Example~\ref{ex:ex1}.
Using either of the feasible matrices $P$ specified in (\ref{eq:p1}) or (\ref{eq:p2}) for the purpose of evaluating the bound (\ref{eq:dwtb_loose_achievable_condition_13}), Theorem~\ref{theo:achstep2} tells us that an a priori unknown feasible location matrix and corresponding value vector can be found to allow perfect recovery of the signal ensemble $X$, as long as $M_1 \ge 2$, $M_2 \ge 2$, and $M_1 + M_2 \ge 5$.
For example, $x_1$ and $x_2$ can be recovered when $M_1= 3$ and $M_2 = 2$.
For this choice of $M_1$ and $M_2$ and for the location matrix $P$ specified in (\ref{eq:p1}), Figure~\ref{fig:matchex} shows that there exists a matching that associates each element of the value vector $\Theta$ to a unique measurement.
Our exposition of the graph based formulation (see Section~\ref{sec:centralpf}) explains how the existence of such a matching ensures perfect recovery of $\Theta$, given $P$, and our proof of Theorem~\ref{theo:achstep2} (see Appendix~\ref{app:achstep2}) explains how the remaining measurements can be used to identify a feasible location matrix.
\begin{figure}[bt]
\begin{center}
\includegraphics[height=5.5cm]{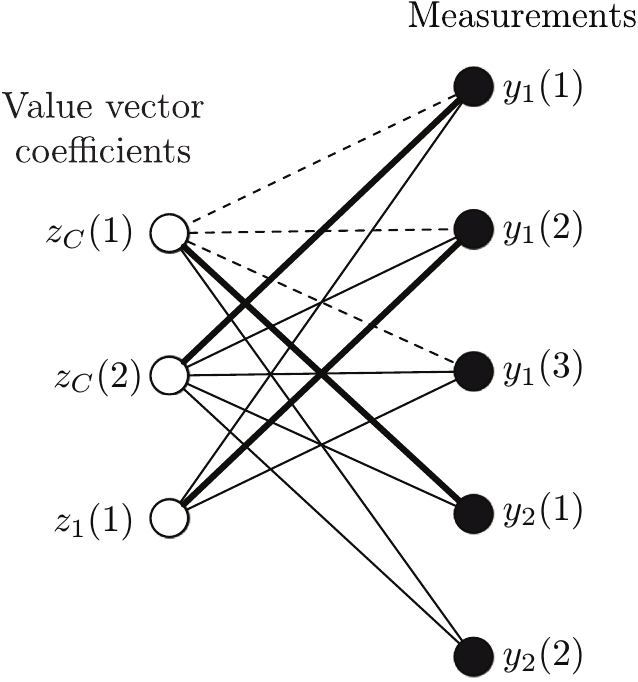}
\end{center}\vspace*{-2mm}
\caption{\small\sl \label{fig:matchex} Graphical representation of the dependencies between value vector coefficients and compressive measurements for the signal ensemble $X$ and location matrix $P$ discussed in Example~\ref{ex:ex4}.
Each edge in this graph denotes a dependency of a measurement on a value vector coefficient, but dashed lines indicate dependencies that cannot be exploited due to overlap of common and innovation coefficients.
Among the edges that remain, the thick solid lines indicate the existence of a matching from each value vector coefficient to a distinct measurement; the existence of such a matching ensures that the system of equations $Y = \Phi P \Theta$ is invertible (see Theorem~\ref{theo:achstep1} and its proof in Section~\ref{sec:centralpf}).
Measurements that remain unassigned in this matching can then be used to verify the correctness of the solution (see Theorem~\ref{theo:achstep2} and its proof in Appendix~\ref{app:achstep2}). }
\vspace{-6mm}
\end{figure}
%
%
\label{ex:ex4}
\end{EXAMPLE}

\section{Central Proof and Bipartite Graph Formulation}
\label{sec:centralpf}

This section is dedicated to proving Theorem~\ref{theo:achstep1}. In order to prove this theorem, we introduce a bipartite graph formulation that represents the dependencies between the obtained measurements $Y$ and the coefficients in the value vector $\Theta$.

\subsection{Proof of Theorem~\ref{theo:achstep1}}
\label{app:achstep1}

For brevity, we denote $K_C(P)$ and $K_j(P)$ simply as $K_C$ and $K_j$. We denote the number of columns of $P$ by
\begin{equation}
\numcols = \numcols(P) := K_C + \sum_{j \in \Lambda} K_j,
\label{eq:numcols}
\end{equation}
and note that $D' \ge D$.
Because $P \in \mathcal{P}_F(X)$, there exists $\Theta \in
\real^\numcols$ such that $X = P \Theta$. Because $Y = \Phi X$,
$\Theta$ is a solution to $Y = \Phi P \Theta$.

We will argue that,
with probability one over $\Phi$, $\Upsilon := \Phi P$ has rank
$\numcols$, and thus $\Theta$ is the unique solution to the equation
$Y =\Phi P \Theta =  \Upsilon \Theta$.
%
To prove that $\Upsilon$ has rank $\numcols$, we invoke the following lemma, which we prove in Section~\ref{app:matching}.

\begin{LEMM}
If (\ref{eq:achCondition1}) holds, then there exists a mapping $\mathcal{C} : \{1,2,\dots,K_C\} \rightarrow \Lambda$, assigning each element of the
common component to one of the sensors, such that for all nonempty subsets $\Gamma \subseteq \Lambda$,
\begin{equation}
\sum_{j \in \Gamma} M_j \;\ge\; \sum_{j \in \Gamma} (K_j + C_j), 
\label{eq:matching}
\end{equation}
where $C_j := | \{ k \in \{1,2,\dots,K_C\}: ~ \mathcal{C}(k) = j \} |$ for each $j \in \Lambda$, and
such that for each $k \in \{1,2,\dots,K_C\}$, the
$k^\mathrm{th}$ column of $P_C$ is not a column of
$P_{\mathcal{C}(k)}$.

\label{lemma:matching}
\end{LEMM}

Intuitively, the existence of such a mapping suggests that ($i$)~each sensor has taken enough measurements to cover its own innovation component (requiring $K_j$ measurements) and perhaps some of the common component, ($ii$)~for any $\Gamma \subseteq \Lambda$, the sensors in $\Gamma$ have collectively taken enough extra measurements to cover the requisite $K_C(\Gamma,P)$ elements of the common component, and ($iii$)~the extra measurements are taken at sensors where the common and innovation components do not overlap.
Formally, we will use the existence of such a mapping to prove that
$\Upsilon$ has rank $\numcols$.

We proceed by noting that $\Upsilon$ has the block structure illustrated in (\ref{eq:pphi}), where each $\Phi_j P_C$ (respectively, $\Phi_j P_j$) is an $M_j \times K_C$ (respectively, $M_j \times K_j$) submatrix of $\Phi_j$ obtained by selecting columns from $\Phi_j$ according to the columns contained in $P_C$ (respectively, $P_j$).
Referring to (\ref{eq:numcols}), we see that, in total, $\Upsilon$ has $\numcols$ columns. To argue that $\Upsilon$ has rank $\numcols$, we will consider a sequence of three matrices $\Upsilon_0$, $\Upsilon_1$, and $\Upsilon_2$ constructed from modifications to $\Upsilon$.

{\bf Construction of $\Upsilon_0$}: We begin by letting $\Upsilon_0$
denote the ``partially zeroed'' matrix obtained from $\Upsilon$ using
the following construction:
%
\begin{enumerate}
\item Let $\Upsilon_0 = \Upsilon$ and $k = 1$.
\item For each $j$ such that $P_j$ has a column that matches column
$k$ of $P_C$ (note that by Lemma~\ref{lemma:matching} this cannot
happen if $\mathcal{C}(k) = j$), let $k'$ represent the column index
of the full matrix $P$ where this column of $P_j$ occurs. Subtract
column $k'$ of $\Upsilon_0$ from column $k$ of $\Upsilon_0$. This
forces to zero all entries of $\Upsilon_0$ formerly corresponding to
column $k$ of the block $\Phi_j P_C$.
\item If $k < K_C$, then increment $k$ and go to step 2.
\end{enumerate}
The matrix $\Upsilon_0$ is identical to $\Upsilon$ everywhere except
on the first $K_C$ columns, where any portion of a column
equal to a column of $\Phi_j P_j$ to its right has been set
to zero.\footnote{We later show (in property \textsf{P3}) that with probability one, none of the columns
become entirely zero.}
Thus, $\Upsilon_0$ satisfies the next two properties, which
will be inherited by matrices $\Upsilon_1$ and $\Upsilon_2$ that we
subsequently define:
\begin{enumerate}
\item[\textsf{P1.}] Each entry of $\Upsilon_0$ is either zero or a Gaussian random
variable.
\item[\textsf{P2.}] All Gaussian random variables in $\Upsilon_0$ are i.i.d.
\end{enumerate}
Finally, because $\Upsilon_0$ was constructed only by subtracting
columns of $\Upsilon$ from one another, $\rank{\Upsilon_0} = \rank{\Upsilon}$.

{\bf Construction of $\Upsilon_1$}: We now let $\Upsilon_1$ be the
matrix obtained from $\Upsilon_0$ using the following construction:
For each $j\in\Lambda$, we select $K_j + C_j$ arbitrary rows from the portion of
$\Upsilon_0$ corresponding to sensor $j$ (the first $M_1$ rows of $\Upsilon_0$ correspond to sensor $1$, the following $M_2$ rows correspond to sensor $2$, and so on).
The resulting matrix $\Upsilon_1$ has
$$
\sum_{j \in \Lambda} (K_j + C_j ) \;=\; \left(\sum_{j \in \Lambda}
K_j\right) + K_C \; = \; \numcols
$$
rows; note that this is fewer than the number of rows in $\Upsilon_0$ if $K_j + C_j < M_j$ for any $j$. Also, because $\Upsilon_1$ was obtained by selecting a subset of rows from $\Upsilon_0$, it has $\numcols$ columns (just like $\Upsilon_0$) and satisfies $\rank{\Upsilon_1} \le \rank{\Upsilon_0} = \rank{\Upsilon}$.

{\bf Construction of $\Upsilon_2$}: We now let $\Upsilon_2$ be the $\numcols \times \numcols$ matrix
obtained by permuting columns of $\Upsilon_1$ using the following
construction:
\begin{enumerate}
\item Let $\Upsilon_2 = [~]$, and let $j = 1$.
\item For each $k$ such that $\mathcal{C}(k) = j$, let
$\Upsilon_1(k)$ denote the $k^\mathrm{th}$ column of $\Upsilon_1$,
and concatenate $\Upsilon_1(k)$ to $\Upsilon_2$, i.e., let
$\Upsilon_2 \leftarrow [\Upsilon_2 ~ \Upsilon_1(k)]$. There are
$C_j$ such columns.
\item Let $\Upsilon_{1,j}$ denote the
columns of $\Upsilon_1$ corresponding to the entries of $\Phi_j P_j$ (the
innovation components of sensor $j$), and concatenate $\Upsilon_{1,j}$
to $\Upsilon_2$, i.e., let $\Upsilon_2 \leftarrow [\Upsilon_2 ~
\Upsilon_{1,j}]$. There are $K_j$ such columns.
\item If $j < J$, then increment $j$ and go to Step 2.
\end{enumerate}
In total, Step 2 chooses $\sum_{j=1}^J C_j = K_C$
columns, while Step 3 chooses $\sum_{j=1}^J K_j$ columns, and thus referring to (\ref{eq:numcols}),
$\Upsilon_2$ has $K_C + \sum_{j=1}^J K_j = D'$ columns. The number of
rows is the same as that of $\Upsilon_1$, making $\Upsilon_2$ a square matrix.
Because $\Upsilon_1$ and $\Upsilon_2$ share the same columns up to
reordering, it follows that
\begin{equation}
\rank{\Upsilon_2} = \rank{\Upsilon_1} \le \rank{\Upsilon}.
\label{eq:rank3}
\end{equation}
Based on its dependency on $\Upsilon_0$, and following from
Lemma~\ref{lemma:matching}, $\Upsilon_2$ meets
properties \textsf{P1} and \textsf{P2} defined above in addition to
a third property:
\begin{enumerate}
\item[\textsf{P3.}] All entries along the main diagonal of $\Upsilon_2$ are Gaussian random
variables (none are deterministically zero).
\end{enumerate}
Property \textsf{P3} follows because each diagonal element of
$\Upsilon_2$ will either be an entry of some $\Phi_j P_j$, which
remains Gaussian throughout our constructions, or it will be an entry
of some $k^\mathrm{th}$ column of some $\Phi_j P_C$ for which
$\mathcal{C}(k) = j$. In the latter case, we know by
Lemma~\ref{lemma:matching} and the construction of $\Upsilon_0$ (Step 2)
that the $k^\mathrm{th}$ column of $\Phi_j P_C$ is not zeroed out, and
thus the corresponding diagonal entry remains Gaussian throughout
our constructions.

Having identified these three properties satisfied by $\Upsilon_2$,
we will prove by induction that, with probability one over $\Phi$,
such a matrix has full rank.

\begin{LEMM}
Let $\Upsilon^{(d-1)}$ be a $(d-1) \times (d-1)$ matrix having full
rank. Construct a $d \times d$ matrix  $\Upsilon^{(d)}$ as follows:
$$
\Upsilon^{(d)} := \left[ \begin{array}{cc} \Upsilon^{(d-1)} & v_1 \\
v_2^T & \omega \end{array} \right]
$$
where $v_1, v_2 \in \real^{d-1}$ are column vectors with each entry being
either zero or a Gaussian random variable, $\omega$ is a Gaussian
random variable, and all random variables are i.i.d.\ and
independent of $\Upsilon^{(d-1)}$. Then with probability one,
$\Upsilon^{(d)}$ has full rank.
\label{lemma:inductionPhi}
\end{LEMM}

Applying Lemma~\ref{lemma:inductionPhi} inductively $\numcols$
times, the success probability remains one. It follows that with
probability one over $\Phi$, $\rank{\Upsilon_2} = \numcols$. Combining
this last result with (\ref{eq:rank3}),
we conclude that $\rank{\Upsilon} = \numcols$ with probability one over $\Phi$.
It remains to prove Lemma~\ref{lemma:inductionPhi}.

\noindent {\bf Proof of Lemma~\ref{lemma:inductionPhi}:} When $d =
1$, $\Upsilon^{(d)} = [\omega]$, which has full rank if and only if
$\omega \neq 0$, which occurs with probability one.

When $d > 1$, using expansion by minors, the determinant of
$\Upsilon^{(d)}$ satisfies
$\det(\Upsilon^{(d)}) = \omega \cdot \det(\Upsilon^{(d-1)}) + C,$
where $C = C(\Upsilon^{(d-1)}, v_1, v_2)$ is independent of
$\omega$. The matrix $\Upsilon^{(d)}$  has full rank if and only if
$\det(\Upsilon^{(d)}) \neq 0$, which is satisfied if and only if
$
\omega \neq \frac{-C}{\det(\Upsilon^{(d-1)})}.
$
By the inductive assumption, $\det(\Upsilon^{(d-1)}) \neq 0$ and $\omega$ is a
Gaussian random variable that is independent of $C$ and
$\det(\Upsilon^{(d-1)})$. Thus, $\omega \neq
\frac{-C}{\det(\Upsilon^{(d-1)})}$ with probability one. \qed

\subsection{Proof of Lemma~\ref{lemma:matching}}
\label{app:matching}

To prove Lemma~\ref{lemma:matching}, we apply tools from graph theory.

\begin{figure}[bt]
\begin{center}
\includegraphics[height=4.5cm]{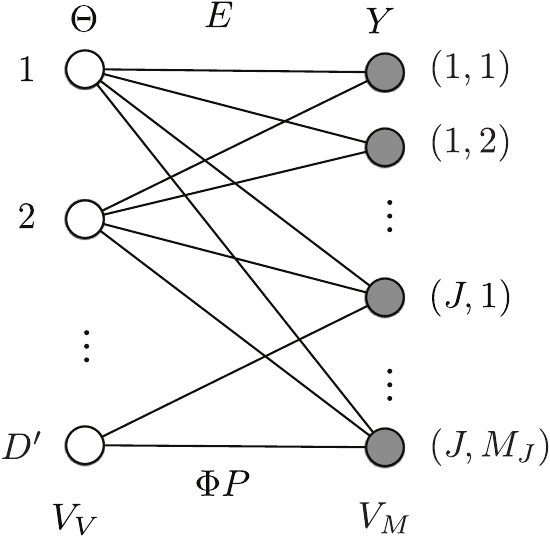}
\end{center}\vspace*{-4mm}
\caption{\small\sl \label{fig:bgraph}
The bipartite graph
$G=(V_V,V_M,E)$ indicates the relationship between the value vector coefficients
and the measurements.}
\vspace{-5mm}
\end{figure}

We introduce a bipartite graph $G=(V_V,V_M,E)$ that captures the dependencies between the entries of the value vector $\Theta \in \real^{D'}$ and the entries of the measurement vector $Y = \Phi P \Theta$.
%
%
This graph is defined as follows. The set of {\em value vertices} $V_V$ has elements with indices $d \in \{1,\ldots,D'\}$ representing the entries $\Theta(d)$ of the value vector.
The set of {\em measurement vertices}
$V_M$ has elements with indices $(j,m)$ representing the measurements
$y_j(m)$, with $j\in\Lambda$ and $m \in \{1,\ldots,M_j\}$ (the range of possible $m$ varies depending on $j$). The cardinalities
for these sets are $|V_V|=D'$ and $|V_M|=\sum_{j\in\Lambda} M_j$.
Finally, the set of edges $E$ is defined according to the following rules:
\begin{itemize}
\item For every $d \in \{1,2,\dots,K_C\} \subseteq V_V$ and $j \in \Lambda$ such
that column $d$ of $P_C$ does not also appear as a column of $P_j$,
we have an edge connecting $d$ to each vertex $(j,m)\in V_M$ for $1
\le m \le M_j$.
\item For every $d \in \{K_C+1, K_C+2,\dots, \numcols\} \subseteq V_V$, we
consider the sensor $j$ associated with column $d$ of $P$, and we
have an edge connecting $d$ to each vertex $(j,m) \in V_M$ for $1 \le m \le M_j$.
\end{itemize}
An example graph for a distributed sensing setting appears in Figure \ref{fig:bgraph}.

We seek a {\em matching} within the bipartite graph $G=(V_V,V_M,E)$, namely, a subgraph $(V_V,V_M,\Etilde)$ with $\Etilde \subseteq \Ebar$ that pairs each element of $V_V$ with a unique element of $V_M$. Such a matching will immediately give us the desired mapping $\mathcal{C}$ as follows: for each $k \in \{1,2,\dots,K_C\} \subseteq V_V$, let $(j,m) \in V_M$ denote the single vertex matched to $k$ by an edge in $\Etilde$; we then set $\mathcal{C}(k) = j$.

To prove the existence of such a matching within the graph, we
invoke a version of Hall's marriage theorem for bipartite
graphs~\cite{graphs}.
Hall's theorem states that within a bipartite graph $(V_1,V_2,E)$,
there exists a matching that assigns each element of $V_1$ to a
unique element of $V_2$ if for any collection of elements $\Pi
\subseteq V_1$, the set $E(\Pi)$ of neighbors of $\Pi$ in $V_2$ has
cardinality $|E(\Pi)| \ge |\Pi|$.
To apply Hall's theorem in the context of our lemma, we will show that if (\ref{eq:achCondition1}) is satisfied, then
for any set $\Pi \subseteq V_V$ of entries in the value vector, the set $\Ebar(\Pi)$ of neighbors of $\Pi$ in $V_M$ has size $|\Ebar(\Pi)| \ge |\Pi|$.

Let us consider an arbitrary set $\Pi \subseteq V_V$. We let  $S_\Pi = \{j \in \Lambda: (j,m) \in \Ebar(\Pi)
\mathrm{~for~some~} m\} \subseteq \Lambda$ denote the set of signal indices
whose measurement vertices have edges that connect to $\Pi$.
Since a connection between a value vertex and a measurement vertex
at a given sensor implies a connection to all other measurement vertices
for that sensor, it follows that $|\Ebar(\Pi)| = \sum_{j \in S_\Pi} M_j$.
Thus, in order to satisfy Hall's condition for $\Pi$, we require
\begin{equation}
\sum_{j \in S_\Pi} M_j \ge |\Pi|. \label{eq:partialhall222}
\end{equation}
We would now like to show that $\sum_{j \in S_\Pi} K_j +K_C(S_\Pi,P) \ge
|\Pi|$, and thus if (\ref{eq:achCondition1}) is satisfied for all
$\Gamma \subseteq \Lambda$, then (\ref{eq:partialhall222}) is
satisfied in particular for $S_\Pi \subseteq \Lambda$.

In general, the set $\Pi$ may contain vertices for both common
components and innovation components. We write $\Pi = \Pi_C \cup
\Pi_I$ to denote the disjoint union of these two sets.

By construction, $|\Pi_I| \le \sum_{j \in S_\Pi} K_j$, because $\Pi_I$ cannot include any innovation component outside the set of sensors $S_\Pi$.
We will also argue that $|\Pi_C| \le K_C(S_\Pi,P)$ as follows.
By definition, for a set $\Gamma \subseteq \Lambda$,
$K_C(\Gamma,P)$ counts the number of columns in $P_C$ that also
appear in $P_j$ for all $j \notin \Gamma$. 
By construction, for each $k \in \Pi_C$, vertex $k$ has no connection to vertices $(j,m)$ for $j
\notin S_\Pi$, and so it must follow that the $k^\mathrm{th}$ column
of $P_C$ is present in $P_j$ for all $j \notin S_\Pi$.  Thus, the index $k$ is among the indices counted in the definition (\ref{eq:kintersect}) of $K_C(S_\Pi,P)$, and therefore $|\Pi_C| \le K_C(S_\Pi,P)$.

We conclude that $|\Pi| = |\Pi_I| + |\Pi_C| \le \sum_{j \in S_\Pi} K_j +K_C(S_\Pi,P)$,
and so (\ref{eq:achCondition1}) implies (\ref{eq:partialhall222})
for any $\Pi$, and so Hall's condition is satisfied, and a matching
exists.
%
%
Finally, consider any set $\Gamma \subseteq \Lambda$.
To confirm that (\ref{eq:matching}) holds for this set, note that there are a total of $\sum_{j \in \Gamma} M_j$ vertices $(j,m) \in V_M$ such that $j \in \Gamma$.
Each of these vertices is matched to at most one vertex in $V_V$, which must correspond either to an innovation component counted in $K_j$ for some $j\in \Gamma$ or to a common component indexed by some $k$ such that $\mathcal{C}(k) \in \Gamma$.
It follows that $\sum_{j \in \Gamma} M_j \ge \sum_{j \in \Gamma} (K_j + C_j)$. \qed

\section{Discussion}
\label{sec-discussion}

In this paper, we have introduced the ensemble sparsity model (ESM) framework for modeling intra- and inter-signal \newtext{dependencies} within a collection of sparse signals.
This framework is based on a factored representation of the signal ensemble that decouples its location information from its value information.
We have also proposed an analytical framework based on bipartite graphs that allowed us, in the context of a common/innovation ESM $\mathcal{P}$, to characterize the numbers of measurements $M_1, M_2, \dots, M_J$ needed for successful recovery of a signal ensemble $X$.
Our bounds highlight the benefit of joint reconstruction in distributed compressive sensing (DCS), since sparse signals can be recovered from fewer measurements than their nominal sparsity level would indicate.

The factored representation that we have proposed for modeling sparse signal ensembles is closely related to the recently proposed union-of-subspaces
modeling frameworks for CS~\cite{dosamplingunion,samplingunion,EldarUSS,modelcs}.
What is particularly novel about our treatment is the explicit consideration of the block structure of matrices such as $P$ and $\Phi$, and the explicit accounting for measurement bounds on a sensor-by-sensor basis.
Most of the conventional union-of-subspaces theory in CS is intended to characterize the number of measurements required to recover a vector $X$ from measurements $\Phi X$, where $\Phi$ is a dense matrix.

Finally, as we have discussed in Section~\ref{sec:ciesm}, common/innovation ESMs can be populated using various choices of matrices $P$ of the form (\ref{eq:pmatrix}).
Our bounds in Section~\ref{sec-theory} are relatively agnostic to such design choices.
Past experience, however, has indicated that practical algorithms for signal recovery can benefit from being tuned to the particular type of signal \newtext{dependencies} under consideration~\cite{DCSTR06,DCSAllerton05,DCSAsilomar05,DCSNIPS,DCSIPSN06,jsconst,EldarMMV,EldarRauhut,BergFriedlander,HoldEm,LeeMUSIC,KimMUSIC,DaviesMUSIC,dirty}. %
Our focus in this paper has been not on tractable recovery algorithms or robustness issues, but rather on foundational limits governing how the measurements may be amortized across the sensors while preserving the information required to uniquely identify the sparse signal ensemble.
However, we believe that our results and our graphical modeling framework may pave the way for a better, perhaps more unified, development of practical DCS algorithms.

\appendices

\section{Proof of Theorem~\ref{theo:cnv}}
\label{app:cnv}

As in (\ref{eq:numcols}), we let $\numcols$ denote the number of columns in $P$. Because $P \in \mathcal{P}_F(X)$, there exists
$\Theta \in \real^\numcols$ such that $X = P \Theta$.
Because $Y = \Phi X$, then $\Theta$ is a solution to $Y = \Phi P
\Theta$.
We will argue for $\Upsilon := \Phi P$ that $\rank{\Upsilon} <
\numcols$, and thus there exists $\widehat{\Theta} \neq \Theta$ such
that $Y = \Upsilon \Theta = \Upsilon \widehat{\Theta}$.
Since $P$ has full rank, it follows that $\widehat{X} := P
\widehat{\Theta} \neq P \Theta = X$.

We let $\Upsilon_0$ be the ``partially zeroed'' matrix obtained from
$\Upsilon$ using the identical procedure detailed in
Section~\ref{app:achstep1}.
Again, because $\Upsilon_0$ was constructed only by subtracting
columns of $\Upsilon$ from one another, it follows that
$\rank{\Upsilon_0} = \rank{\Upsilon}$.

Suppose that $\Gamma \subseteq \Lambda$ is a set for which
(\ref{eq:cnvCondition1}) holds.
We let $\Upsilon_3$ be the submatrix of $\Upsilon_0$ obtained by
selecting the following columns:
\begin{itemize}
\item For any $k \in \{1,2,\dots,K_C\}$ such that column $k$ of
$P_C$ also appears as a column in $P_j$ for all $j \notin \Gamma$,
we include column $k$ of $\Upsilon_0$ as a column in $\Upsilon_3$.
There are $K_C(\Gamma,P)$ such columns $k$. 
\item For any $k \in \{K_C+1,K_C+2,\dots,\numcols\}$ such that
column $k$ of $P$ corresponds to an innovation for some sensor $j
\in \Gamma$, we include column $k$ of $\Upsilon_0$ as a column in
$\Upsilon_3$. There are $\sum_{j \in \Gamma} K_j$ such columns $k$.
\end{itemize}
This submatrix has
$\sum_{j \in \Gamma} K_j+K_C(\Gamma,P)$
columns. Because $\Upsilon_0$ has the same size as $\Upsilon$
(see Section~\ref{app:achstep1}), and
in particular has only $\numcols$ columns, then in order to have
that $\rank{\Upsilon_0} = \numcols$, it is necessary that all
$\sum_{j \in \Gamma} K_j+K_C(\Gamma,P)$ columns of $\Upsilon_3$ be
linearly independent.

Based on the method described for constructing
$\Upsilon_0$, it follows that $\Upsilon_3$ is zero for all
measurement rows not corresponding to the set $\Gamma$.
These rows were nonzero only for two sets of columns of $\Upsilon_0$:
($i$) the columns corresponding to the innovations for signals
$j \notin \Gamma$, and ($ii$) the columns $k \in \{1,2,\ldots,K_C\}$ for
which the $k^{th}$ column of $P_C$ appears in none of the matrices $P_j, j \notin \Gamma$.
Both of these sets of columns are discarded during the construction of $\Upsilon_3$.
Therefore, consider the submatrix $\Upsilon_4$ of $\Upsilon_3$ obtained
by selecting only the measurement rows corresponding to the set
$\Gamma$. Because all the rows discarded from $\Upsilon_3$ are zero,
it follows that $\rank{\Upsilon_3} = \rank{\Upsilon_4}$. However, since
$\Upsilon_4$ has only $\sum_{j \in \Gamma} M_j$ rows, we invoke
(\ref{eq:cnvCondition1}) and have that $\rank{\Upsilon_3} =
\rank{\Upsilon_4} \le \sum_{j \in \Gamma} M_j <
\sum_{j \in \Gamma} K_j+K_C(\Gamma,P)$. Thus, all
$\sum_{j \in \Gamma} K_j+K_C(\Gamma,P)$ columns of $\Upsilon_3$ cannot be
linearly independent, and so $\Upsilon$ does not have full rank.
This means that there exists $\widehat{\Theta} \neq \Theta$ such
that $Y = \Upsilon \Theta = \Upsilon \widehat{\Theta}$, and thus
we cannot distinguish between the two solutions
$\widehat{X} := P \widehat{\Theta} \neq P \Theta = X$.
\qed

\section{Proof of Theorem~\ref{theo:achstep2}}
\label{app:achstep2}

Given the measurements $Y$ and measurement matrix $\Phi$, we will
show that it is possible to recover some $P\in
\mathcal{P}_F(X)$ and a corresponding vector $\Theta$ such that
$X = P \Theta$ using the following algorithm.
\begin{itemize}
\item Extract from each measurement vector $y_j$ its final entry, and sum these entries to obtain the quantity $\overline{y} = \sum_{j \in \Gamma} y_j(M_j)$. Similarly, add the corresponding rows of $\Phi$ into a single row $\overline{\phi}^T$. The row vector $\overline{\phi}^T$ is a concatenation of the final rows of the matrices $\Phi_j$, and thus its entries are i.i.d.\ Gaussian. Note that $\overline{y} = \overline{\phi}^T X$; this quantity will be used in a cross-validation step below.
\item Group the remaining $\left(\sum_{j\in \Lambda} M_j\right) - J $ measurements into a vector $\overline{Y}$, and let $\overline{\Phi}$ contain the corresponding rows of $\Phi$. We note that $\overline{\phi}$ is independent from $\overline{\Phi}$ and that $\overline{Y} = \overline{\Phi} X$.
\item For each matrix $P \in \mathcal{P}$ such that $\overline{Y} \in \mathrm{colspan}(\overline{\Phi} P)$, choose a single solution $\Theta_P$ to $\overline{Y} = \overline{\Phi} P \Theta_P$ independently of $\overline{\phi}$. Then, perform the following cross-validation: if $\overline{y} = \overline{\phi}^T P \Theta_P$, then return the estimate $\widehat{X} = P\Theta_P$; otherwise, continue with the next
    matrix $P$.
\end{itemize}
We begin by noting that there exists at least one matrix $P \in \mathcal{P}$ for which $\overline{Y} \in \mathrm{colspan}(\overline{\Phi} P)$ and for which $X = P \Theta_P$.
In particular, consider the matrix $P^* \in \mathcal{P}_F(X)$ mentioned in the theorem statement.
Because (\ref{eq:dwtb_loose_achievable_condition_13}) holds for $P^*$, Theorem~\ref{theo:achstep1} guarantees that with probability one, $\overline{\Phi} P^*$ will have full rank, and so there is a unique solution $\Theta_{P^*}$ to $\overline{Y} = \overline{\Phi} P^* \Theta_{P^*}$.
Since $P^* \in \mathcal{P}_F(X)$ and $P^*$ is full rank, we know that $X = P^* \Theta_{P^*}$.
Also, since $Y = \Phi X$, we know that $\overline{y} = \overline{\phi}^T P^* \Theta_{P^*}$, and so this matrix will clear the cross-validation step.

Now suppose that, for some $P \in \mathcal{P}$, the algorithm above considers a candidate solution $\Theta_P$ to $\overline{Y} = \overline{\Phi} P \Theta_P$, but suppose also that $X \neq P \Theta_P$.
The algorithm will fail to discard this incorrect solution if $\Theta_P$ passes the cross-validation test, i.e., if $\overline{\phi}^T P \Theta_{P}=\overline{y} = \overline{\phi}^T X$.
Recall, however, that $\overline{\phi}$ is an i.i.d.\ Gaussian random vector and that it is independent of both $X$ and $P \Theta_P$.
It then follows that $\overline{\phi}$ is orthogonal to $X - P \Theta_P$ with probability zero, and therefore we will have $\overline{\phi}^T (X - P \Theta_P) \neq 0$ (equivalently, $\overline{\phi}^T P \Theta_{P} \neq \overline{y}$) with probability one.
Therefore, this incorrect solution will be discarded with probability one.
Since $\mathcal{P}$ contains only a finite number of matrices, the probability of cross-validation discarding all incorrect solutions remains one. \qed

\vspace{-2mm}
\section*{Acknowledgments}
\vspace{-1mm}

Thanks to Emmanuel Cand\`{e}s, Albert Cohen, Ron DeVore, Anna Gilbert, Illya Hicks, Robert Nowak, Jared Tanner, and Joel Tropp for informative and inspiring conversations.
%
%
Portions of this work were performed at Princeton University and Duke University, the University of Michigan, and the Technion by MFD, MBW, and DB, respectively.
The authors thank these institutions (and in particular Prof.\ Tsachy Weissman) for their hospitality.

\begin{IEEEbiographynophoto}
{Marco F. Duarte}
(S'99--M'09) received the B.Sc. degree in computer engineering (with 
distinction) and the M.Sc. degree in electrical engineering from the 
University of Wisconsin-Madison in 2002 and 2004, respectively, and the 
Ph.D. degree in electrical engineering from Rice University, Houston, TX, in 
2009. He was an NSF/IPAM Mathematical Sciences Postdoctoral Research 
Fellow in the Program of Applied and Computational Mathematics at 
Princeton University, Princeton, NJ, from 2009 to 2010, and in the 
Department of Computer Science at Duke University, Durham, NC, from 2010 
to 2011. He is currently an Assistant Professor in the Department of Electrical 
and Computer Engineering at the University of Massachusetts, Amherst, MA. 
His research interests include machine learning, compressed sensing, sensor 
networks, and optical image coding.\\
Dr. Duarte received the Presidential Fellowship and the Texas Instruments 
Distinguished Fellowship in 2004 and the Hershel M. Rich Invention Award 
in 2007, all from Rice University. He is also a member of Tau Beta Pi.
\end{IEEEbiographynophoto}

\begin{IEEEbiographynophoto}
{Michael B. Wakin} 
(S'01--M'07) received the B.A.\ degree in mathematics in 2000, and the B.Sc., M.Sc., and Ph.D.\ degrees, all in electrical engineering, all from Rice University, Houston, TX, in 2000, 2002, and 2007, respectively. He was an NSF Mathematical Sciences Postdoctoral Research Fellow with the California Institute of Technology, Pasadena, from 2006 to 2007, and an Assistant Professor with the University of Michigan, Ann Arbor, from 2007 to 2008. He is currently an Assistant Professor with the Department of Electrical Engineering and Computer Science at the Colorado School of Mines. His research interests include sparse, geometric, and manifold-based models for signal and image processing, approximation, compression, compressive sensing, and dimensionality reduction\\
Dr.\ Wakin shared the Hershel M. Rich Invention Award in 2007 from Rice University for the design of a single-pixel camera based on compressive sensing; in 2008, he received the DARPA Young Faculty Award for his research in compressive multisignal processing; and in 2012, he received the NSF CAREER Award for research into dimensionality reduction techniques for structured data sets.
\end{IEEEbiographynophoto}

\begin{IEEEbiographynophoto}
{Dror Baron} 
(M'03--SM'10) received the B.Sc.\ (summa cum laude) and M.Sc.\ degrees from the Technion---Israel Institute of Technology, Haifa, in 1997 and 1999, and the Ph.D.\ degree from the University of Illinois at Urbana-Champaign in 2003, all in electrical engineering.
From 1997 to 1999, he was with Witcom Ltd., working in modem design. From 1999 to 2003, he was a Research Assistant at the University of Illinois at Urbana-Champaign, where he was also a Visiting Assistant Professor in 2003. He
was a Postdoctoral Research Associate in the Department of Electrical and Computer Engineering at Rice University, Houston, TX, from 2003 to 2006. He was a quantitative financial analyst with Menta Capital, San Francisco, CA, from 2007 to 2008, and a Visiting Scientist in the Department of Electrical Engineering at the Technion---Israel Institute of Technology, Haifa, from 2008 to 2010. He is currently an Assistant Professor in the Electrical and Computer Engineering Department at the North Carolina State University. His research interests combine information theory, signal processing, and fast algorithms; in recent years, he has focused on compressed sensing.\\
Dr. Baron was a recipient of the 2002 M. E. Van Valkenburg Graduate Research Award, and received honorable mention at the Robert Bohrer Memorial Student Workshop in April 2002, both at the University of Illinois. He also participated from 1994 to 1997 in the Program for Outstanding Students, comprising the top 0.5\% of undergraduates at the Technion.
\end{IEEEbiographynophoto}

\begin{IEEEbiographynophoto}
{Shriram Sarvotham} 
(S'00--M'01) received the B.Tech.\ degree from the Indian Institute of Technology, Madras, India, and the M.S.\ and Ph.D.\ degrees from Rice University, Houston, TX, all in electrical engineering. His research interests lie in the broad areas of compressed sensing, nonasymptotic information theory, and Internet traffic analysis and modeling. Currently, he works as a Principal Research Scientist at Halliburton Energy Services, Houston, TX, where he investigates optimal data acquisition and processing of NMR data in oil and gas exploration.
\end{IEEEbiographynophoto}

\begin{IEEEbiographynophoto}
{Richard G. Baraniuk} 
(M'93--SM'98--F'01) received the B.Sc.\ degree in 1987 from the
University of Manitoba (Canada), the M.Sc.\ degree in 1988 from the
University of Wisconsin-Madison, and the Ph.D.\ degree in 1992 from the
University of Illinois at Urbana-Champaign, all in Electrical Engineering.
After spending 1992--1993 with the Signal Processing Laboratory of
Ecole Normale Sup\'{e}rieure, in Lyon, France, he joined Rice
University, where he is currently the Victor E.\ Cameron Professor
of Electrical and Computer Engineering.   His research interests lie in 
the areas of signal processing, machine learning, and open education.

Dr.\ Baraniuk received a NATO postdoctoral fellowship from NSERC in
1992, the National Young Investigator award from the National
Science Foundation in 1994, a Young Investigator Award from the
Office of Naval Research in 1995, the Rosenbaum Fellowship from the
Isaac Newton Institute of Cambridge University in 1998, the  C.\
Holmes MacDonald National Outstanding Teaching Award from Eta Kappa
Nu in 1999, the University of Illinois ECE Young Alumni
Achievement Award in 2000, the Tech Museum Laureate Award from the 
Tech Museum of Innovation in 2006, 
the Wavelet Pioneer Award from SPIE in 2008, 
the Internet Pioneer Award from the Berkman Center
for Internet and Society at Harvard Law School in 2008, the
World Technology Network Education Award and IEEE Signal Processing 
Society Magazine Column Award in 2009, the IEEE-SPS
Education Award in 2010, the WISE Education Award in 2011, 
and the SPIE Compressive Sampling Pioneer Award in 2012.
In 2007, he was selected as one of 
Edutopia Magazine's Daring Dozen educators, and
the Rice single-pixel compressive camera was selected by MIT Technology
Review Magazine as a TR10 Top 10 Emerging Technology. 
He was elected a Fellow of the IEEE in 2001 and of AAAS in 2009.
\end{IEEEbiographynophoto}
\end{document}